# Linear and nonlinear fiber propagation of partially coherent fields exhibiting temporal correlations

*Frédéric. Audo,* [1] *Hervé Rigneault,*[2] *and Christophe Finot* [1,*]

1 Laboratoire Interdisciplinaire Carnot de Bourgogne, UMR 6303 CNRS - Université de Bourgogne-Franche-Comté, 9 avenue Alain Savary, BP 47870, 21078 Dijon Cedex, France

2 Aix Marseille Université, CNRS, Centrale Marseille, Institut Fresnel UMR 7249, 13397

E-mail: christophe.finot@u-bourgogne.fr

Using ultrafast photonic first-order differentiator applied on a partially coherent field, we report the generation of two correlated temporal waveforms and study their correlation properties upon linear and nonlinear propagation along the two orthogonal polarization axis of a dispersive optical fiber. Temporal correlations are maintained in linear propagation whereas Kerr nonlinearity generates anti-correlated temporal intensity patterns for both partially and uncorrelated fields. Experiments are in close agreement with the theoretical analysis.

## 1. Introduction

The powerful analogy between the spatial and temporal evolutions of optical waves has stimulated a large number of recent researches.[1-4] Indeed, the temporal consequences of dispersion are formally identical to the spatial evolution of a 1D beam affected by diffraction. Given this powerful duality, numerous papers have reported the temporal analogs of well-known optical systems processing coherent signals such as lenses,[2] lenticular lenses,[5] dispersion grating,[6] interference devices [7] and recently the famous Arago spot formation.[8] Processing of partially coherent signals has also been explored.[1] However, at this stage, the number of studies remains rather limited with examples of temporal imaging [9-11] or ghost imaging.[12] On the contrary, the spatial domain has attracted much more interest regarding the diffraction of partially coherent fields, especially with the studies dealing with speckle and scattering in random media.[13-15] As a recent example, it has been shown that the statistical properties of the speckle could be easily manipulated taking advantage of singularities found in speckle fields.[16] This was done using spiral phase patterns imprinted on a wavefront prior to its propagation in a random medium. It was shown that spiral phases ($\pi$ phase shifts between two opposite points in the pupil plane) of different handiness resulted in complementary speckle intensity patterns (where maxima and minima are exchanged).[16]

In the present contribution, we propose to study a similar situation in the time domain by means of a Fourier processing that can generate two stochastics fields with complementary temporal properties (where maxima and minima are exchanged). We experimentally demonstrate that performing a temporal derivation (through an all optical first-order differentiator) we can efficiently transform a partially coherent field to generate complementary patterns in the time domain. Using a polarization multiplexing scheme, we experimentally confirm that such complementary features can be maintained over dispersive propagation in fiber. We also discuss experimentally and numerically the impact of Kerr non-linearity (self- and cross-phase modulation) on the temporal complementary of partially coherent and incoherent fields and find that Kerr processes may deeply affect the temporal field correlations.

## 2. Principle and model

### 2.1. Input fields

We consider a partially coherent optical field typical of a spectrally filtered amplified spontaneous emission whose spatial structure is assumed to be a plane-wave. In the approximation of the slowly variable envelope, this initial random field *u(t)* has a Fourier transform $u(\omega)$ which is Gaussian-shaped with $\delta$-correlated random spectral phase $\varphi(\omega)$ uniformly distributed between $-\pi$ and $\pi$ :[17]

$$u(\omega) \propto \exp\left(-2\ln(2)\frac{\omega^2}{\Omega^2}\right) \exp(i\,\varphi(\omega)). \tag{1}$$

with $\Omega$ the full-width at half maximum (fwhm) spectral width of the intensity profile that is typically $2\pi \times 20$ Grad.s$^{-1}$ in the experiment we present below in section 3. In the temporal domain (see **Figure 1**(a), blue line), this corresponds to large and ultrashort fluctuations of the intensity profile $U(t) = u^2$, with a minimum fwhm temporal width of $T = 0.44 / (\Omega/2\pi) = 22$ ps. As the spectrum is $\delta$-correlated, the wave exhibits intensity fluctuations that are statistically stationary in time. An indirect but convenient mean to have an idea of the temporal fluctuations is to use intensity correlations defined for two temporal intensity profiles $F(t)$ and $G(t)$ as :

$$Corr_{F,G}(\tau) = \int F(t)\, G(t+\tau)\, dt. \tag{2}$$

Using a normalized autocorrelation function $\Gamma_U = [Corr_{U,U}(\tau) - Corr_{U,U}(\tau=\infty)]/[Corr_{U,U}(0) - Corr_{U,U}(\tau=\infty)]$, we obtain that $\Gamma_U$ is a Gaussian function with a fwhm temporal width being $\sqrt{2}\,T$ (see Figure 1b, blue line):

$$\Gamma_U(\tau) = \exp\left(-2\ln(2)\frac{\tau^2}{T^2}\right). \tag{3}$$

In ref [16], a spiral phase mask was inserted in the Fourier plane of a lens in order to deterministically modify the spatial properties of the speckle under study and to create complementary speckle patterns (as compared to the generated without the spiral phase mask). The temporal analogue would be to perform a Hilbert transform, i.e. to insert a $\pi$ phase shift between frequencies located below and above the pulse central frequency. From the space-time duality it is expected that such Hilbert transform would create a complementary light intensity time profile compared to the temporal intensity profile of the initial pulse (i.e. without performing the Hilbert transform). However, given the limited resolution of optical spectral shapers, an ideal spectral phase jump associated with an all-pass response is impossible to achieve. A π phase shift is in practice associated with a change of the spectral intensity profile such as a notch filtering response at the center of the spectrum. Instead of implementing a first-order photonic Hilbert transformer and given the interrelations of the two processes,[18, 19] we chose here to perform a first-order time derivative (first-order differentiator). Such a function is optically and easily achieved in the spectral domain by multiplying the incident field by $i\,\omega$, i.e. by applying a quadratic transfer function on the spectral intensity profile (x $\omega^2$) combined with a π-phase shift at the pulse central frequency ($\omega = 0$). As a consequence, the spectrum obtained after temporal optical derivation becomes:

$$v(\omega) \propto \sqrt{\omega^2 \exp\left(-4\ln(2)\frac{\omega^2}{\Omega^2}\right)} \exp\left[i\,\varphi(\omega) + \pi\,H(-\omega)\right]. \qquad (4)$$

with $H(\omega)$ the Heaviside step function.

An example of the resulting temporal intensity profile $V(t)$ is provided in Figure 1(a) (red curve). We can note a good match between the maxima of the initial intensity profile $U$ and the instants when $V$ vanishes. The zeros of $U$ are also often associated with maxima of $V$. Those two observations are however not fully systematic. Indeed, the optical first-order differentiator is coherent and affects the complex fields $u$ and $v$ and not directly $U$ and $V$. However, our intuitive expectation, that is here numerically confirmed, is that, as the intensity temporal maxima also correspond to region of slowly varying phase, the intensity maxima and zeros of $U$ and $V$ are most of the times interchanged. We also plotted in green in Figure 1(a) the signal made by the incoherent sum of the two intensities $U+V$. Such a signal presents typical fluctuations that appear longer than the fluctuations of $U$ or $V$.

In order to better quantify such temporal fluctuations, we have computed the auto-correlations of $V$ and $U+V$ from an average procedure made over an ensemble of 500 numerical simulations over a duration of 1 ns. Results (red and green lines) are plotted in Figure 1(b) and compared with the properties of the input field (blue line). It is also possible to derive analytically the autocorrelation properties using the Wiener-Khintchine theorem and we note that the analytical results (circles) are in perfect agreement with the numerical simulations. In the case of field resulting from the optical derivation, $\Gamma_V$ is provided by :

$$\Gamma_V(\tau) = \Gamma_U(\tau) \left[1 - 2\ln(2)\frac{\tau^2}{T^2}\right]^2. \tag{5}$$

This autocorrelation signal presents a shape that differs significantly from the initial field, with the existence of pronounced sidelobes and a central part that is narrower. Regarding the wave resulting from the incoherent superposition of the two complementary patterns, the autocorrelation signal of $U+V$ has an extended coherence with respect to $U$ or $V$ (with a fwhm of the normalized autocorrelation that is 33% wider). The complementarity between the two signals $U$ and $V$ is also apparent in the normalized cross-correlation of the signal $U$ and $V$ defined as $C_{U,V} = [Corr_{U,V}(\tau) - Corr_{U,V}(0)]/(\sigma_U \sigma_V)$ ($\sigma_U$ and $\sigma_V$ denoting the standard deviations of $U$ and $V$) that can be analytically computed as :

$$C_{U,V} = 2\ln(2)\frac{\tau^2}{T^2}\Gamma_U(\tau). \tag{6}$$

This cross-correlation (Figure 1c) vanishes for null delays confirming that the bright temporal spots of $U$ do not heavily overlap with those of $V$. The maxima of the cross-correlation are obtained for $\tau = T/\sqrt{2\ln(2)}$, which can be interpreted as a repulsion delay between $U$ and $V$. Note that as $\Gamma_{U+V}(\tau) = (\Gamma_U(\tau) + \Gamma_V(\tau) + 2\,C_{U,V}(\tau))/2$, $\Gamma_{U+V}$ can be derived analytically from the knowledge of Equation (3), (5) and (6) :

$$\Gamma_{U+V}(\tau) = \Gamma_U(\tau)\left[1 + 2\left(\ln 2\,\frac{\tau^2}{T^2}\right)^2\right]. \tag{7}$$

## 2.2. Linear and nonlinear field propagation:

We concentrate now on the linear and nonlinear propagation of the correlated fields *u* and *v*. In this context, single-mode optical fibers are excellent testbeds, enabling the propagation over long distances without any change in the spatial beam profile. One of the experimental challenges is to propagate both incoherent signals over the same fiber length with the same linear and nonlinear properties and to separate them at the fiber output for characterization. A way to meet this requirement is to benefit from fiber polarization multiplexing. Using polarization beam combiner and splitter, *u* and *v* can propagate on two orthogonal polarizations within the same fiber. The longitudinal evolution of the complex slowly varying amplitudes *u(z,t)* and *v(z,t)* are then described by a set of two coupled nonlinear Schrödinger (NLS) equations that take into account the dispersion ($\beta_2$ being the group velocity dispersion coefficient) and the instantaneous Kerr nonlinearity ($\gamma$ being the nonlinear Kerr parameter of the fiber). In standard optical fibers, averaging out the nonlinear contribution over the fast random polarization fluctuations in km-long fibers leads to the so-called Manakov model:[20, 21]

$$\begin{cases} i\dfrac{\partial u}{\partial z} + \dfrac{\beta_2}{2}\dfrac{\partial^2 u}{\partial t^2} + \dfrac{8}{9}\gamma\left(|u|^2 + |v|^2\right)u + i\dfrac{\alpha}{2}u = 0, \\ i\dfrac{\partial v}{\partial z} + \dfrac{\beta_2}{2}\dfrac{\partial^2 v}{\partial t^2} + \dfrac{8}{9}\gamma\left(|u|^2 + |v|^2\right)v + i\dfrac{\alpha}{2}v = 0 \end{cases}. \qquad (8)$$

where *z* and *t* denote the propagation distance coordinate and time in the co-moving frame of the waves. $\alpha$ stands for the linear losses of the fiber. Equations (8) can be solved thanks to the split-step Fourier method.[22] More advanced approaches to handle the NLS equation exist based on the wave-turbulence, kinetics description and thermalization of optical waves, but they are well beyond the scope of the present discussion.[23, 24] When dispersion is the main effect (i.e. when the average power of the transmitted field is negligible), dispersion leads to the development of a quadratic spectral phase that does not modify the power spectrum. In this context, auto- and cross- correlations functions should not be affected. When the average powers of the waves become more significant, the picture gets more complex and self- and cross- phase modulation impairs the propagation and may change the overall power spectrum. The statistical properties of the intensity profiles *U* and *V* may, therefore, be affected, as well as their mutual coherence. As an example of the changes that may occur, in ref. [25], two mutually incoherent signals change over propagation to ultimately generate polarization domain walls. Note that recently, the study of a strong coherent field on a weak signal probe has been discussed for a similar experimental configuration.[26] In the next section, we implement an experimental framework to study the linear and nonlinear propagation of the random fields *u* and *v* in optical fiber.

## 3. Experimental setup

We developed an all-fibered experimental setup relying on commercially available devices used for the telecommunication industry (**Figure 2**). An amplified spontaneous emission source (ASE) operating in the conventional band of optical communication (1550 nm) is linearly polarized before being spectrally tailored by a programmable LCOS device (waveshaper finisar) [27, 28] that can handle the simultaneous shaping of two different polarization states. On one polarization, the reference beam *u*, the spectral shaper imprints an optical bandpass filter (OBPF) $f_u(\omega)$ with a Gaussian shape having a fwhm spectral width of 20 GHz. On the second polarization, depending on the configuration under test, the same OBPF shape $f_v = f_u$ may be applied (leading to identical signal and intensity profiles *u(t)= v(t)* and *U(t)= V(t)*), or the filter $f_v$ under test may correspond to the transfer function typical of the optical first-order differentiator defined by $f_v = f_u\, F(\omega)$ with $F(\omega) = i\,\omega$. Note that similar optical processing could also be achieved using a long-period fiber grating,[29] a fiber Bragg grating,[30] a conventional two-arm interferometer [31] or with on-chip microdisk resonators.[19] Given the large spectral bandwidth of the ASE (over 1 THz) which also exhibits a high spectral flatness, it is not needed to take into account the fine details of the ASE profile. Also note that by removing the polarizer after the ASE source and $f_v = f_u$, it is possible to investigate the case of two fully uncorrelated fields having the same power spectrum (in this case the new signal and intensity profiles are noted $u'(t)$ and $U'(t)$). The polarization controller inserted before the programmable pulse shaper is used to equalize the average powers on the two output polarizations of the shaper. Whereas in the spatial domain,[16] the characterization of the speckle grains can be achieved thanks to a set of images recorded successively when applying the Fourier spiral phase, the problem in our case is a bit trickier. Indeed, the non-repetitive nature of the stochastic ps fluctuations generated by the ASE source requires to record simultaneously and synchronously the two ultrafast signals $U(t)$ and $V(t)$. The delay between $u$ and $v$ injection has been carefully adjusted so as to be as low as possible at the input of the fiber. In order to synchronize the two fields and therefore compensate for a slight mismatch in the component lengths used for $u$ and $v$, we have taken advantage of the spectral shaper to imprint an additional linear spectral phase (delay).

$u(t)$ and $v(t)$ are amplified thanks to an erbium-doped fiber amplifier (EDFA) able to deliver average powers up to 23 dBm. For such average powers and given the spectral bandwidths of the signals, it is possible to neglect higher-order linear or nonlinear effects such as third-order dispersion, Raman or Brillouin scattering. The propagation can, therefore, be accurately described by the Manakov equations given in Equation (8).

The dispersive propagation occurs in a dispersion compensation fiber (DCF) with a length of 13 km and a strong normal dispersion of $\beta_2 = 0.1$ ps$^2$/m. The nonlinear evolution is studied in a dispersion-shifted fiber (DSF) with a length of 10 km having a normal dispersion of 18.2 ps$^2$/km and a nonlinear coefficient of 2.2 /W/km. The linear losses are .2 dB/km. After the propagation in the fiber, the polarization controller and a polarization beam splitter enable us to demultiplex $u$ and $v$. The crosstalk between the two channels is around -20 dB.

The temporal intensity profiles *U* and *V* are simultaneously detected thanks to two high-speed photodiodes with similar properties and having a detection bandwidth of 50 GHz at -3 dB. A variable attenuation is used to maintain the same average power on the photodiodes, whatever the power delivered by the EDFA. A real-time oscilloscope with a bandwidth of 50 GHz enables to save 500 recordings of 1 ns that are then used to numerically compute the auto- and cross-correlations signals. Note that those signals could also be recorded directly using optical auto- and cross-correlators based on second-harmonic generation or on two-photon absorption in a semiconductor. The spectral content of the signal is monitored using a high-resolution optical spectrum analyzer.

## 4. Experimental results

### 4.1. Generation of the various correlated waves

Taking advantage of the linear shaper, we have tailored the spectral properties of the partially coherent field coming from the ASE source. The experimental spectral intensity profile of the reference beam *u* is compared in **Figure 3** with the profile obtained after the optical first-order differentiation. Both profiles reproduce with high accuracy the theoretical shape that is programmed on the optical shaper.

**Figure 4**(a) illustrates the temporal intensity profiles of the signals *U(t)* and *V(t)* recorded on both polarization components. The main features previously discussed and illustrated in Figure 1(a) are clearly reproduced experimentally: intensity maxima and nearly-zero regions of partially coherent fields are experimentally interchanged. As an additional consequence, the maxima of *U* and *V* are clearly not synchronized. We can also note that the computed sum of the two signals (green curve) appears broader than the two signals taken independently. The resulting auto-correlation signals are provided in Figure 4(b). Slight deviations from the expected theoretical shape are visible. Indeed, whereas $\Gamma_v$ should reach values down to zero

between the sidelobes, we can note that such a value is not reached experimentally. This mismatch is largely ascribed to the finite bandwidth of our detection bandwidth, as shown by the much better agreement obtained between the experimental data and the numerical results taking into account the finite bandwidth of the photodiodes and oscilloscope as well as a slight crosstalk between the two polarization channels. Experimental autocorrelation of $U+V$ confirms that the temporal coherence region of the superposition of the two waves is increased compared to $U$ alone.

The cross-correlation signals that are recorded also confirm that our device is able to synthesize partially coherent fields with the expected mutual coherence properties. Therefore, when the polarizer in front of the programmable filter is removed, the resulting random signals obtained on both polarizations are completely uncorrelated as shown by the coefficient $C_{U,U'}$ close to zero for any value (black curve). When the polarizer is inserted and the same filter is applied to the $u$ and $v$ signals, we retrieve $C_{U,U} = \Gamma_U$ and a correlation coefficient that reaches 1 for $\tau = 0$. In the case of the first-order differentiation which is the point of this study, we observe that the correlation factor severely drops in the central part of the cross-correlation signal $C_{U,V}$. The value of the maximum of $C_{U,V}$ as well as its temporal location are in close agreement with the theoretical expectations. However, contrary to the theoretical expectations, the value of the correlation does not fully vanish for a null delay. The wings of the cross-correlation function are also slightly broader than theoretically expected. Once again, these discrepancies are essentially ascribed to the limit of our detection in terms of bandwidth and crosstalk. Numerically taking into account these limitations enables us to reproduce the experimental results (Figure 4c, blue circles).

**4.2. Propagation in a linear medium**

We investigate as a first step the ability of the tailored stochastic wave to maintain their correlation properties upon dispersive propagation. We consider here an optical fiber with a high level of dispersion that induces an integrated dispersion of 1 300 ps$^2$ upon propagation over 13 km. In such a fiber, a fully coherent pulse of duration $T$ = 22 ps would experience a significant broadening up to 150 ps. However, for a partially incoherent and stationary signal, we do not observe such a broadening in the autocorrelation measurements. As shown in **Figure 5**, the correlation properties of the signal are not affected by the linear propagation [22] as it is confirmed by the near identical records of the auto- and cross- correlation signals measured at the input and output of the fiber. We have also checked that the optical spectra of *u* and *v* (not shown here) do not present any change.

### 4.3. Nonlinear propagation of the signal

We now investigate the impact of nonlinearity on the incoherent light propagation. Self-phase modulation is known to impact the optical spectrum of optical waves.[24, 32, 33] In order to quantify how the properties of the incoherent waves are affected, we carried out a series of measurements according to the input average power. The experimental evolution of the power spectrum as a function of the input power is summarized on panels (a) of **Figure 6**. We can note the changes of the optical spectra, with a growth of the spectral tails. Regarding the spectrum of *v*, we can also notice that the notch-filter structure that is preserved for linear propagation tends to be progressively filled and to disappear. This spectral evolution is closely reproduced by numerical simulations of Equation (8) presented in panels (b).

Given the interaction with dispersion, these changes of the power spectrum impact the temporal properties of the waveforms. As shown in panels (c) of Figure 6, the autocorrelations records are strongly affected by the nonlinearity with the developments of large temporal pedestals.

Once again, the experiments are in agreement with the numerical simulations when the limitations of our temporal detection are included into the modeling.

One important question that arises is the way the cross-correlation between $U$ and $V$ is affected by the presence of Kerr nonlinearity. In order to quantify this evolution, we focus on the value of $C_{U,V}$ at null delay. Results of numerical simulations from Equation (8) for different levels of input powers are summarized in **Figure 7**. These numerical simulations show that contrary to linear evolution, $C_{UV}(\tau=0)$ is not maintained when nonlinearity impairs propagation. When cross-phase modulation is not taken into account (i.e. the evolution of $u$ and $v$ are not coupled in Equation (8)), the value of $C_{U,V}$ at $\tau = 0$ tends to increase with power (green curve). This strongly contrasts with $C_{U,V}(\tau = 0)$ evolution when cross-phase modulation links the evolutions of $u$ and $v$ : when XPM is taken into account, the value of $C_{U,V}(\tau = 0)$ decreases and reaches negative values (blue). Such negative values of $C_{U,V}$ indicate an anticorrelation: at times where $U$ increases, $V$ decreases, and vice versa. This evolution of the orthogonal polarization components can be linked to the results reported by Gilles *et al* [25] that have demonstrated the generation of polarization domains walls starting from two incoherent and non-correlated orthogonally polarized fields. It is therefore of interest to compare the evolution obtained from our two stochastic but correlated waves with what can be achieved starting from uncorrelated waveforms (red curve). Quite unexpectedly, it appears that for the present range of parameters, both initial conditions (correlated and uncorrelated) tend to provide a similar evolution, confirming the fact that the decreases of the $C_{U,V}$ at $\tau = 0$ is governed by a robust process of nonlinear attraction rather than a process dictated by the initial conditions.

Results of the experimental records of the evolution of the cross-correlation $C_{U,V}$ are reported in **Figure 8** for correlated and uncorrelated fields. In agreement with the numerical simulations, we observe that both correlated and uncorrelated conditions lead to similar trends, with a decrease of $C_{U,V}(\tau = 0)$ and negative values that can be reached with increasing power.

However, given the finite bandwidth of our detection a discrepancy in the value of $C_{U,,V}(\tau = 0)$ is visible between the correlated and uncorrelated cases. Such a discrepancy can be fully reproduced by numerical simulations when including the limitations of the optoelectronics detection. The full temporal records of $C_{U,,V}$ are also provided in panel (b) and confirm that experimentally, a dip appears and increases for $C_{U,,V}(\tau = 0)$. Once again, full numerical simulations that take into account the detection response provide a good agreement with experiments.

## 5. Conclusion

Several conclusions can be drawn from this work. We have experimentally demonstrated that temporal photonic first-order differentiation can be extended beyond the bounds of coherent processes. Similarly to its spatial counterpart,[16] this ultrafast treatment of a partially coherent field leads to a temporal complementary arrangement of the maxima and null values of the field. Using polarization multiplexing in an optical fiber, we have shown that these particular correlation properties can be maintained over significant distances under linear propagation. When nonlinear cross- and self-phase modulations affect the evolution in a normally dispersive fiber, the dynamics becomes more complex and the two partially coherent fields asymptotically tend to the formation of polarization domain walls with an anticorrelated nature. We captured the experimental evidence of a nonlinear self-organization process among the orthogonal polarization components that does not seem to depend on the initial correlation properties of the incoherent waves that are involved. We also stress the excellent agreement between the experimental measurements and the numerical simulations including experimental limitations of our detection setup.

This proof-of-concept experiment therefore paves the way for more complex Fourier processing of stochastic fields. Indeed, ultrafast analog processing such as higher-order optical differentiation or integration can also be implemented.[34] Our fiber-based setup appears as a convenient platform to study the linear or nonlinear evolution of two correlated signals. We got here interested in intensity correlations, but one can also plan to investigate field correlations. Future studies could also be dedicated to clarify how the proposed process is able to impact the probability distribution functions for the resulting partially coherent fields.[35]

**Acknowledgements**
We thank Julien Fatome for fruitful discussions. This work has benefited from the PICASSO experimental platform of the University of Burgundy. It was funded by the Agence Nationale de la Recherche (ANR) (ANR-11-LABX-01-01), the Région Bourgogne-Franche-Comté (PARI Photcom), the Program FEDER-FSE Bourgogne 2014-2020 and the Institut Universitaire de France (IUF).

References

[1]　V. Torres-Company, J. Lancis, P. Andrés, Chapter 1 - Space-Time Analogies in Optics, in: E. Wolf (Ed.) Progress in Optics, Elsevier2011, pp. 1-80.

[2]　B.H. Kolner, *IEEE J. Quantum Electron.* **1994**, *30,* 1951-1963.

[3]　R. Salem, M.A. Foster, A.L. Gaeta, *Adv. Opt. Photon.* **2013**, *5,* 274-317.

[4]　L. Froehly, F. Courvoisier, D. Brunner, L. Larger, F. Devaux, E. Lantz, J.M. Dudley, M. Jacquot, *J. Opt. Soc. Am. A* **2019**, *36,* C69-C77.

[5]　J. Nuno, C. Finot, J. Fatome, *Opt. Fiber Technol.* **2017**, *36,* 125-129.

[6]　C. Finot, H. Rigneault, *J. Opt. Soc. Am. B* **2017**, *34,* 1511-1517.

[7]　F. Chaussard, H. Rigneault, C. Finot, *Opt. Commun.* **2017**, *397,* 31-38.

[8]     C. Finot, H. Rigneault, *J. Opt* **2019**, *21,* 105504.

[9]     B. Li, J. Azaña, *IEEE Photon. J.* **2015**, *7,* 1-8.

[10]    B. Li, J. Azaña, *J. Lightw. Technol.* **2016**, *34,* 2758-2773.

[11]    P. Naulleau, E. Leith, *Appl. Opt.* **1995**, *34,* 4119-4128.

[12]    P. Ryczkowski, M. Barbier, A.T. Friberg, J.M. Dudley, G. Genty, *Nat. Photon.* **2016**, *10,* 167.

[13]    G. Gbur, T.D. Visser, Chapter 5 - The Structure of Partially Coherent Fields, in: E. Wolf (Ed.) Progress in Optics, Elsevier2010, pp. 285-341.

[14]    M. Born, E. Wolf, Principles of Optics, Seventh Edition ed., Cambridge University Press1999.

[15]    T.D. Visser, G.P. Agrawal, P.W. Milonni, *Opt. Lett.* **2017**, *42,* 4600-4602.

[16]    J. Gateau, H. Rigneault, M. Guillon, *Phys. Rev. Lett.* **2017**, *118,* 043903.

[17]    J.W. Goodman, Statistical optics, John Wileay and Sons1985.

[18]    N.Q. Ngo, Y. Song, *Opt. Lett.* **2011**, *36,* 915-917.

[19]    T. Yang, J. Dong, L. Liu, S. Liao, S. Tan, L. Shi, D. Gao, X. Zhang, *Sci. Rep.* **2014**, *4,* 3960.

[20]    P.K.A. Wai, C.R. Menyuk, H.H. Chen, *Opt. Lett.* **1991**, *16,* 1231-1233.

[21]    D. Marcuse, C.R. Menyuk, P.K.A. Wai, *J. Lightw. Technol.* **1997**, *15,* 1735-1746.

[22]    G.P. Agrawal, Nonlinear Fiber Optics, Fourth Edition, Academic Press, San Francisco, CA, 2006.

[23]    A. Picozzi, *Opt. Express* **2007**, *15,* 9063-9083.

[24]    P. Suret, A. Picozzi, S. Randoux, *Opt. Express* **2011**, *19,* 17852-17863.

[25]    M. Gilles, P.Y. Bony, J. Garnier, A. Picozzi, M. Guasoni, J. Fatome, *Nat. Photon.* **2017**, *11,* 102.

[26]    J. Nuno, C. Finot, G. Millot, M. Erkintalo, J. Fatome, *Commun. Phys.* **2019**, *2,* 138.

[27]   A.M. Clarke, D.G. Williams, M.A.F. Roelens, B.J. Eggleton, *J. Lightw. Technol.* **2010**, *28,* 97-103.

[28]   J. Dong, Y. Yu, Y. Zhang, B. Luo, T. Yang, X. Zhang, *IEEE Photon. J.* **2011**, *3,* 996-1003.

[29]   R. Slavík, Y. Park, M. Kulishov, R. Morandotti, J. Azaña, *Opt. Express* **2006**, *14,* 10699-10707.

[30]   M.J. Li, D. Janner, J. Yao, V. Pruneri, *Opt. Express* **2009**, *17,* 19798-19807.

[31]   Y. Park, J. Azaña, R. Slavík, *Opt. Lett.* **2007**, *32,* 710-712.

[32]   J.T. Manassah, *Opt. Lett.* **1990**, *15,* 329-331.

[33]   B. Barviau, S. Randoux, P. Suret, *Opt. Lett.* **2006**, *31,* 1696-1698.

[34]   J. Azaña, *IEEE Photon. J.* **2010**, *2,* 359-386.

[35]   N. Bender, H. Yılmaz, Y. Bromberg, H. Cao, *Optica* **2018**, *5,* 595-600.

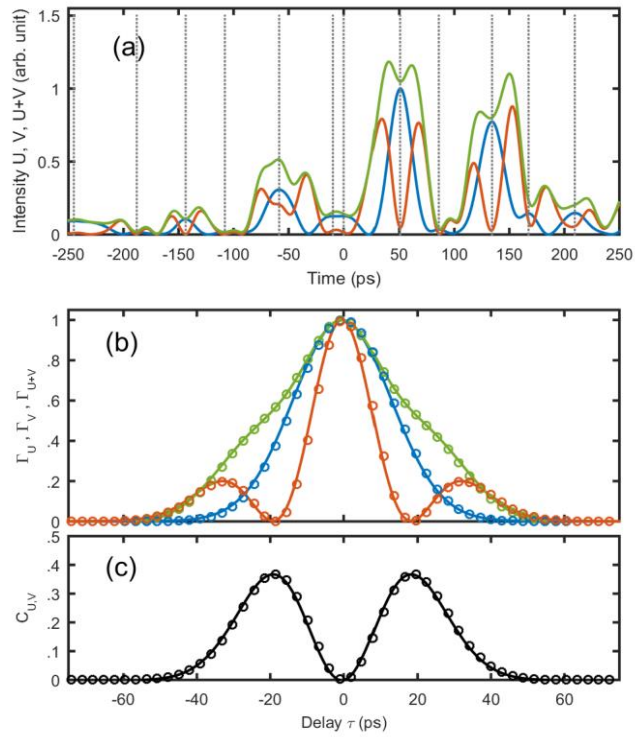

**Figure 1.** (a) Temporal evolution of the intensity profile of the partially incoherent wave *U* (blue line) compared with the wave *V* (red line) and with the sum *U+V* (green line). The vertical lines represent the temporal position of the extrema of *U*. Signals *U* and *V* are normalized so that they have the same average power. (b) Normalized intensity auto-correlations of the signals *U*, *V* and *U+V* (blue, red and green colors, respectively). Numerical simulations (solid line) are compared to the analytical results provided by Equation (3), (5) and (7) (circles). (c) Cross-correlations $C_{U,V}$ between signals *U* and *V*. Numerical simulations (solid line) are compared to the analytical results provided by Equation (6) (circles).

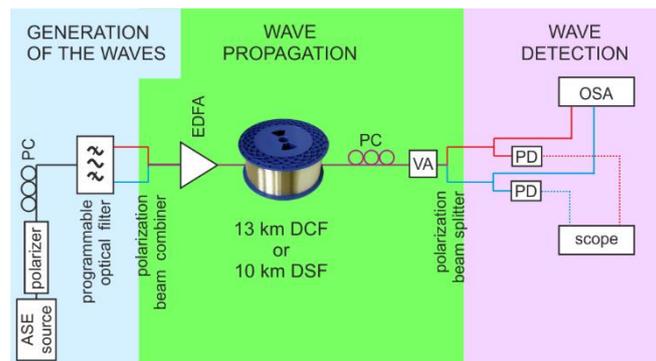

**Figure 2.** Implemented experimental setup. ASE : Amplified Spontaneous Emission source, PC : Polarization Controler, DCF : Dispersion Compensating Fiber, DSF : Dispersion-Shifted Fiber, VA : Variable Attenuator, PD : Photodiode, OSA : Optical Spectrum Analyzer

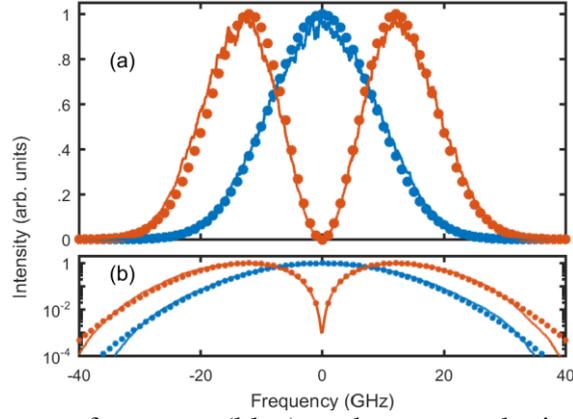

**Figure 3.** Power spectrum of wave *u* (blue) and wave *v* obtained after optical first-order differentiation (red). Experimental results (solid line) are compared with the shaping targets (Equation (1) and (4), circles). Results on (a) a linear scale and (b) logarithmic scale.

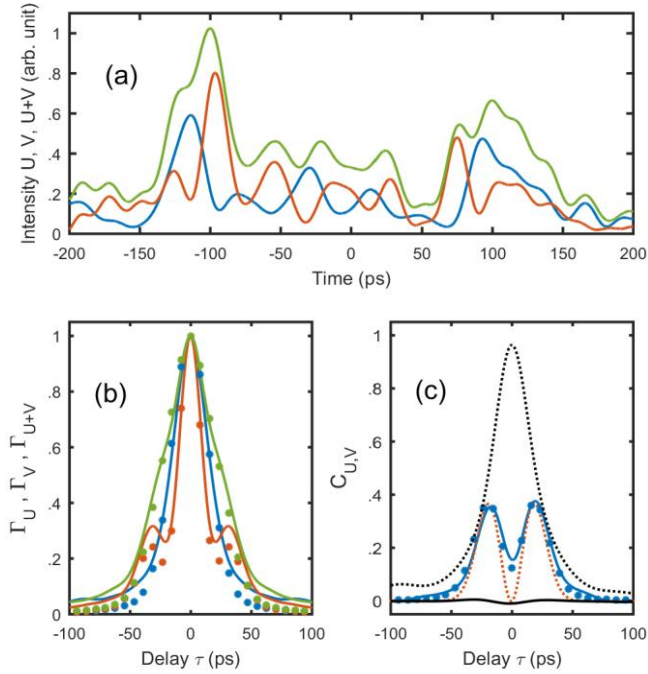

**Figure 4.** Temporal properties of the signals $I_u$, $I_v$ and the sum $I_u + I_v$ (blue, red and green colors, respectively). (a) Temporal intensity profiles of the partially coherent waves recorded experimentally. (b) Auto-correlation profiles. Experimental results (lines) are compared with numerical simulations (circles) taking into account the finite bandwidth and the crosstalk. (c) Cross-correlation measurements in different configurations: when *V = U* (dashed black line), when *V* and *U* are uncorrelated (solid black line) and when *V* results from the optical first-order differentiation of *u* (blue dashed line for experimental results that are compared with the theoretical results (dashed red) and with numerical simulations including experimental limitations (blue circles)).

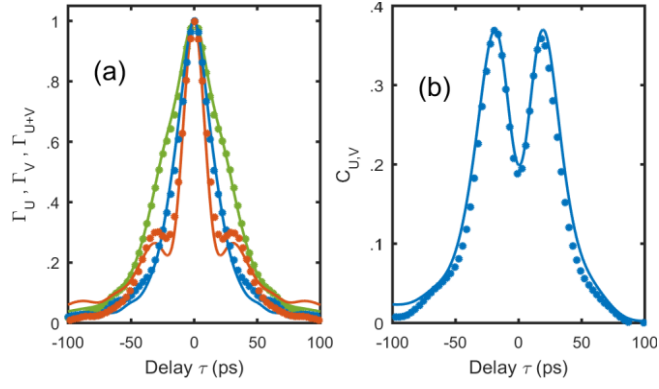

**Figure 5.** Propagation in a dispersive medium of partially incoherent pulses: (a) Comparison of the various autocorrelation signals at the input (dots) and output of the fiber (solid line). Same color code as in Fig 1(b). (b) Cross-correlation signal between *U* and *V*.

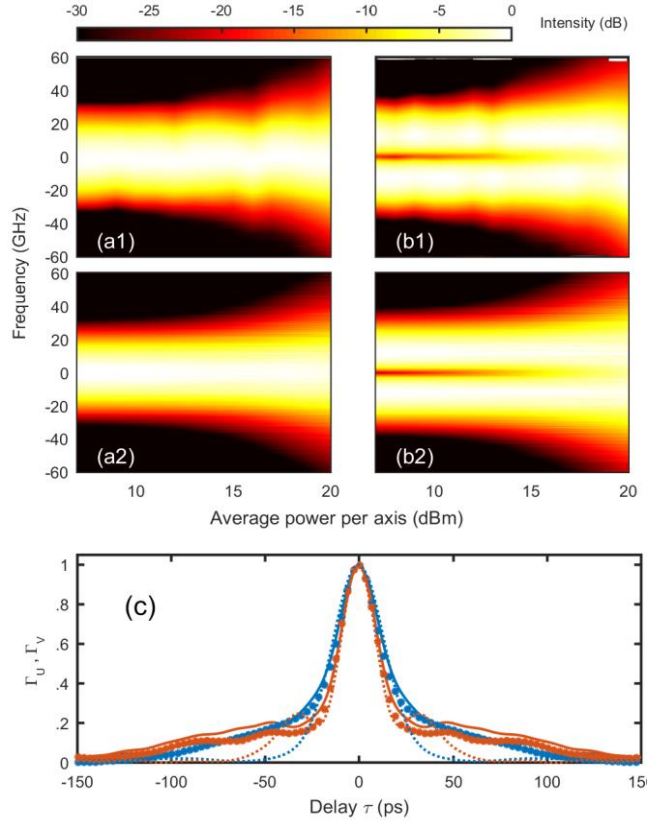

**Figure 6.** Nonlinear propagation of partially coherent waves. (a-b) Optical spectra for the partially coherent wave *u* and the wave obtained after first-order differentiation *v* (panels a and b, respectively) as a function of the average power sent in the fiber per axis. Experimental results (panels 1) are compared with the numerical simulations of Eq. (8) (panels 2). (c) Auto-correlation signal recorded for wave *U* and *V* (blue and red respectively). The experimental autocorrelations obtained after nonlinear propagation (solid line) are compared with their initial values (dotted lines). Results of numerical simulations including detection limitation are overlaid (*U*: blue circles, V: red circles). In (c) results are obtained for an average power of 20 dBm per axis.

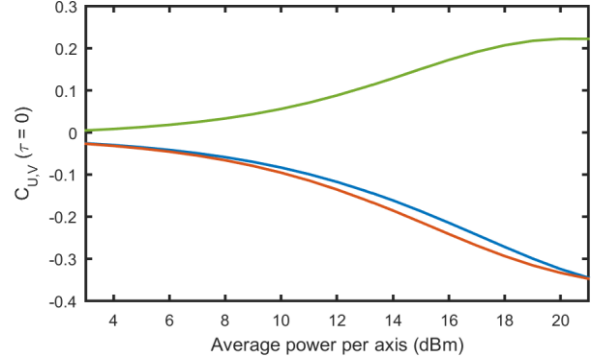

**Figure 7.** Influence of the average power on the cross-correlation $C_{U,V}$ at null delay in the case of the co-propagation of the two waves without XPM (green curve) and with XPM (blue line). The results obtained with two non-correlated fields are also plotted (red solid line). All the results are obtained from numerical simulations.

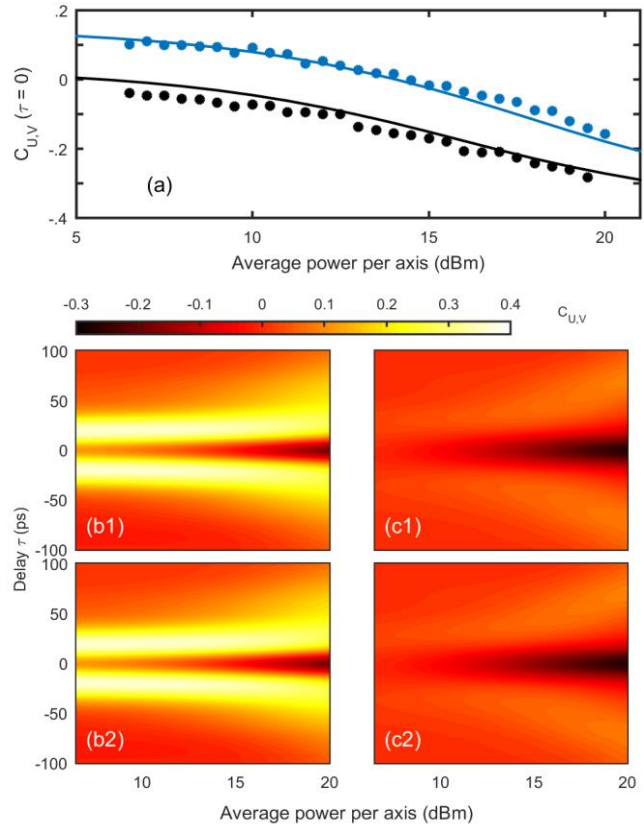

**Figure 8.** Influence of the average power on the cross-correlation between *U* and *V* for correlated and uncorrelated fields. (a) Evolution of the value of the cross-correlation at null delay $C_{U,V}(\tau = 0)$ for the case of two uncorrelated waves (black) and in case of two correlated waves *U* and *V* obtained after first-order differentiation (blue). The results of the numerical simulations including the bandwidth restrictions (solid line) are compared with the experimental results (circles). (b-c) Evolution of the cross-correlation $C_{U,V}$ of the two correlated wave *U* and *V* (b) and uncorrelated waves (c). Experimental results (b1 and c1) are compared with numerical simulations including the detection bandwidth limit (panels b2 and c2).